\documentclass[preprint,showpacs,preprintnumbers,amsmath,amssymb]{revtex4}

\usepackage{graphicx}
\usepackage{dcolumn}
\usepackage{bm}

\begin{document}

\title{Magnetohydrodynamic Stability at a
  Separatrix: Part I} 
\author{A.J. Webster \& C.G. Gimblett}
\affiliation{Euratom/UKAEA Fusion Association, Culham Science Centre,
  Abingdon, Oxfordshire, OX14 3DB.} 
 \email{anthony.webster@ukaea.org.uk}

\date{\today}

\begin{abstract}
The rapid deposition of energy by Edge Localised Modes (ELMs) onto
plasma facing components, is a potentially serious issue for 
large Tokamaks such as ITER and DEMO. 
The trigger for ELMs is believed to be the ideal Magnetohydrodynamic
Peeling-Ballooning instability, but recent numerical calculations have
suggested that a plasma equilibrium with an X-point - as is found in
all ITER-like Tokamaks, is stable to the Peeling mode. 
This contrasts with analytical calculations
(G. Laval, R. Pellat,  J. S. Soule, Phys Fluids, {\bf 17}, 835,
(1974)), that found the Peeling mode to be unstable in cylindrical
plasmas with arbitrary cross-sectional shape.  
However the analytical calculation only applies to a Tokamak
plasma in a cylindrical approximation.
Here, we 
re-examine the assumptions made in cylindrical geometry calculations,
and generalise the calculation to an arbitrary
Tokamak geometry at marginal stability. 
The resulting equations solely describe the Peeling mode, and are not
complicated by coupling to the ballooning mode, for example. 
We find that stability is determined by the value of a single
parameter $\Delta'$ that is the poloidal average of the
normalised jump in the radial derivative of the perturbed magnetic
field's normal component. 
We also find that near a separatrix it is possible for the energy
principle's $\delta W$  
to be negative (that is usually taken to indicate that the mode is
unstable, as in the cylindrical theory), but the growth rate to be
arbitrarily small.


\end{abstract}


\maketitle

\section{Introduction}

Thermonuclear fusion requires plasmas with a pressure of at least an
atmosphere, and temperatures in excess of 100 million degrees Kelvin. 
These conditions can be achieved in Tokamaks such as JET\cite{Wesson},
but  the plasmas are subject to a number of
instabilities, the consequences of which range from benign to
structurally damaging.
By understanding the instabilities that can occur, they can be avoided
or mitigated. 
A class of instabilities that are only partly understood are Edge
Localised Modes (ELMs)\cite{Zohm}.
ELMs can lead to a
rapid deposition of  energy onto plasma facing components, and this 
is a potentially serious issue for proposed large tokamak devices 
such as ITER\cite{Aymar}.  

Our present understanding of ELMs is based on the linear ideal
Magnetohydrodynamic Peeling-Ballooning instability (Wilson et al
\cite{Wilson99}, Gimblett et al\cite{Gimblett}), which is 
thought to trigger ELMs, that subsequently evolve
non-linearly. The studies  upon which this understanding were based 
considered  Tokamak equilibria with a smoothly shaped magnetic
flux-surface at  the plasma-vacuum boundary. In contrast, modern
Tokamak plasmas  have a cross-section in which the outermost flux
surface is redirected onto divertor plates, forming a separatrix with
a sharp ``X-point'' where the magnetic topology changes from closed
(confined plasma), to open field lines along which plasma can flow to
the divertor plates\cite{Wesson}.

The first numerical evidence for a stabilising effect from the
separatrix was found by Medvedev et al\cite{Medvedev}. 
More recently, numerical studies of the Peeling-Ballooning
instability in these X-point plasmas (Huysmans\cite{Huysmans}), have
found that 
as the plasma's  outermost flux surface is made increasingly close to
that of a separatrix  with an X-point, the Peeling mode becomes
stabilised. Crucially the  stabilisation appeared to happen before the
plasma formed  a separatrix with an X-point, shaping alone appeared to
be sufficient  to stabilise the mode.  
This appears to be contrary to theoretical work by Laval et
al\cite{Laval},   
which has indicated that the peeling mode is unstable in cylindrical
plasmas with an arbitrarily shaped cross-section. 
In addition the ELITE code (Wilson et al\cite{Wilson02}) has recently
been used to examine Peeling mode stability  
as the outermost flux surface approaches the separatrix. 
It was found
that although the growth rate reduced in size as the boundary more
closely approximated a separatrix, it did so increasingly slowly, and
its asymptotic behaviour was uncertain (Saarelma\cite{Saarelma}). 
To help understand and reconcile these results, here in the first part
of this two  part paper we re-examine the assumptions made in the
derivation of the 
peeling mode  stability criterion for a cylinder, and generalise the
calculation so that it applies to a toroidal Tokamak plasma.  

As with the original studies of the Peeling mode in a cylinder, for
simplicity we will firstly consider marginal stability, and generalise
the 
condition for Peeling mode stability in a straight cylinder to a
condition for Peeling mode stability in 
an arbitrary cross-section Tokamak plasma. The resulting equations
only describe Peeling mode stability, and are not complicated by
coupling to the Ballooning mode instability, for example. 

This paper generalises previous analytic calculations in a number of
ways. Firstly it applies to axisymmetric toroidal geometries, as
opposed to the cylindrical geometry in which the Peeling mode has been
extensively studied (for example see Laval et al\cite{Laval},
Lortz\cite{Lortz}, Connor et al\cite{Connor}). 
It allows for equilibrium poloidal currents at the plasma
edge, in addition to the toroidal current that is solely included in
previous analytic studies. 
The skin currents that are induced by a plasma perturbation are
related to the difference between the magnetic field in the plasma and
the vacuum, and it is found  
that at marginal stability a plasma perturbation induces a skin
current that is parallel and proportional to the equilibrium edge
current and proportional to the amplitude of the radial plasma
displacement. 
The complicated-looking plasma-vacuum boundary condition that is
usually found in association with the energy principle may be
expressed as a simple relationship between the normal components of
the plasma and vacuum magnetic fields.  
This is used to relate the generalised equations for Peeling mode
stability at marginal stability to the energy principle, and this
allows us to define the Peeling mode in terms of  
the energy principle's $\delta W$. 
With this energy principle for the Peeling mode we can consider the
trial function used by Laval et al\cite{Laval}, finding that a
single parameter $\Delta'$ determines the sign and magnitude of
$\delta W$.  
Finally the instability's growth rate is considered.

\section{Background}

Laval et al\cite{Laval} considered a large aspect ratio ordering  
that neglects toroidal effects, and also neglects any equilibrium 
poloidal current at the plasma edge. 
The work suggests that the Peeling mode will be unstable for a
non-zero edge current, regardless of the plasma cross section. Later
we will reconsider Peeling mode stability for arbitrary cross section
Tokamak plasmas, but firstly we consider some properties of the trial
function considered by Laval et al\cite{Laval}. 

\begin{figure}[tpb!]
\begin{center}
\resizebox{100mm}{!}
{\includegraphics[angle=-90.]{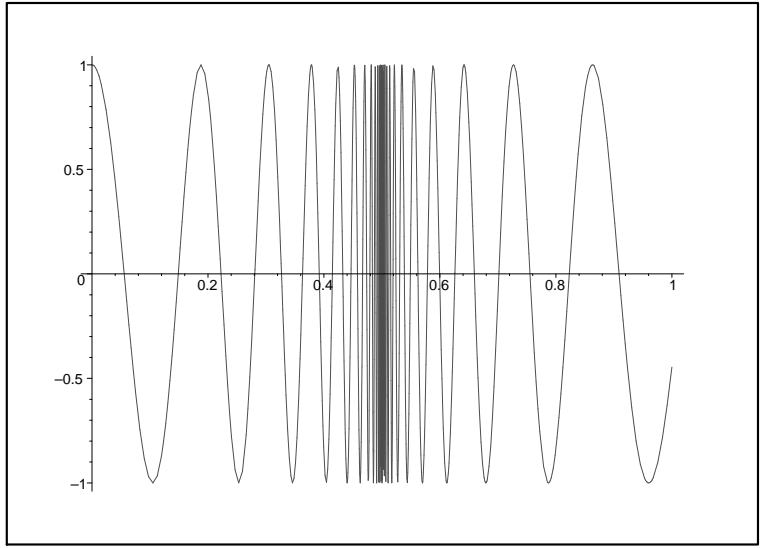}}
\caption{
A plot of the trial function used by Laval et al\cite{Laval},
that for the most unstable modes with $m\sim nq$ has $\xi \sim
e^{inq\theta}$, with $\theta$ the usual straight field line
angle\cite{Bateman}.  
In the second part to this paper we will show that the length $l$
along a flux surface in the poloidal cross section may be
parameterised by $\alpha$, with $\alpha=-\pi..\pi$, as can $\theta$.  
It is also shown that a physically reasonable model for $\theta$ has 
$\theta(\alpha)\simeq \frac{1}{q} \int^{\alpha}_{-\pi}
\frac{d\alpha}{\left( \frac{\alpha^2}{2} + \epsilon^2 \right)}$ and
$l(\alpha)/l(\pi)\simeq \int^{\alpha}_{-\pi} \left( \frac{\alpha^2}{2}
+ \epsilon^2 \right)^{1/4} d\alpha / \oint \left( \frac{\alpha^2}{2} +
\epsilon^2 \right)^{1/4} d\alpha$.  
Therefore by parameterising both $\theta(\alpha)$ and $l(\alpha)$,
then the figure plots $\theta$ (vertical axis) versus $l/l(\pi)$
(horizontal axis), for $\epsilon=0.001$ and  
$n=20$.  
}\label{xiofchi}  
\end{center}
\end{figure}

Note that for an element of length $dl$ along a flux surface in the
poloidal plane, $\frac{dl}{B_p}= \frac{J_{\chi}B_p}{B_p}d\chi=\frac{\nu
  R^2}{I}d\chi$ with $J_{\chi}$ the Jacobian, $B_p$ the poloidal
field, $\nu=\frac{IJ_{\chi}}{R^2}$ is the local field-line pitch, and 
$\chi$,  $\phi$, $\psi$  an orthogonal toroidal co-ordinate
system (for example,  see Freidberg\cite{Freidberg} for details). 
Thus the poloidal angle used in Laval et al, $2\pi \int_0^l
\frac{dl}{B_p} / \oint \frac{dl}{B_p} = 2\pi \int^{\chi}\nu
d\chi'/\oint \nu d\chi' = \theta$ is the same as the usual straight
field line angle\cite{Bateman},  
and $q=\frac{1}{2\pi}\oint \nu d\chi = \frac{1}{2\pi} \oint
\frac{I}{R^2} \frac{dl}{B_p}$.  
The perturbation they consider has a plasma displacement $\xi \sim
e^{im\theta}$, so if we plot $\xi$ versus the length along a flux
surface in the poloidal plane,  then near the 
separatrix in an X-point equilibrium $\xi$ will oscillate arbitrarily
rapidly as we approach the X-point (see figure \ref{xiofchi}). 
The most unstable modes have $m\simeq nq$, so in the figure we plot a
mode with $m\simeq nq$, for which $e^{im\theta} \simeq
e^{in\int^{\chi} \nu d\chi}$. 
Alternately, when $m\ll nq$ or $m \gg nq$, then the mode localises
poloidally in 
the vicinity of the X-point, and is approximately constant elsewhere.  
The rapid oscillation of $\xi$ near the X-point makes it questionable
whether it is physically acceptable, and other terms beyond ideal MHD
need to be considered, but it certainly means that the
closer we approach the separatrix the greater the number of Fourier
modes required (since $m\simeq nq$), and the smaller a computer code's
mesh spacing 
would need to be to represent the mode. 
Therefore as we approach the
separatrix it will be increasingly difficult for a numerical
calculation to represent the mode. 

\section{Cylindrical Plasmas}

Here we outline the derivation of the marginal
stability condition for an arbitrarily large aspect ratio
(cylindrical) equilibrium with the Tokamak
ordering (Freidberg\cite{Freidberg}, Wesson\cite{Wesson}).  
We start from the usual force balance equation $\vec{J}\wedge \vec{B}
= \nabla p$, and take the curl of both sides, expanding to give, 
\begin{equation}
0 = \vec{B}.\nabla \vec{J} - \vec{J} . \nabla \vec{B}
\end{equation} 
Linearising the equation then gives for the
equilibrium quantities
\begin{equation}\label{00}
0 = \vec{B}_0.\nabla \vec{J}_0 - \vec{J}_0 . \nabla \vec{B}_0
\end{equation}
and for the perturbed quantities 
\begin{equation}\label{linearised}
0 = \vec{B}_0 . \nabla \vec{J}_1 + \vec{B}_1 . \nabla \vec{J}_0 
- \vec{J}_0 . \nabla \vec{B}_1 - \vec{J}_1 . \nabla \vec{B}_0 
+\xi . \nabla \left( \vec{B}_0.\nabla \vec{J}_0 - \vec{J}_0 . \nabla
\vec{B}_0    \right)
\end{equation}
Because Eq. \ref{00} holds everywhere, with 
$\vec{B}_0.\nabla \vec{J}_0 - \vec{J}_0 . \nabla \vec{B}_0$ having the
constant value of zero, the last term in Eq. \ref{linearised} that
arises from the  
 displacement of the plasma surface by $\xi$, is zero. 
In the large aspect ratio approximation (Freidberg\cite{Freidberg},
Wesson\cite{Wesson}),  
Eq. \ref{linearised}  
further simplifies to  
\begin{equation}\label{lin2}
0 = \vec{B}_0 . \nabla \vec{J}_1 + \vec{B}_1 . \nabla \vec{J}_0 
\end{equation}

Again, in the large aspect ratio (cylindrical) approximation we may write  
$\vec{B}_1=\vec{e}_z \wedge
\nabla \tilde{\psi}$,  for which 
\begin{equation}
\begin{array}{ll}
\vec{J}_1 &= \nabla \wedge \left( \vec{e}_z \wedge \nabla \tilde{\psi}
\right)  
\\
&= \vec{e}_z \nabla^2 \tilde{\psi} - \vec{e}_z . \nabla \nabla \tilde{\psi} 
\end{array}
\end{equation}
With this same ordering (Freidberg\cite{Freidberg}, Wesson\cite{Wesson}), 
$\vec{J}_0$ is taken to be parallel to $\vec{e}_z$.
Therefore the  $\vec{e}_z$ component of Eq. \ref{lin2}
gives 
\begin{equation}
0= \vec{B}_0 . \nabla \left( \vec{J}_1 . \vec{e}_z \right) + \vec{B}_1
. \nabla J_{0z} 
\end{equation}
Substituting for $\vec{J}_1$ then gives 
\begin{equation}\label{mstab1} 
0= \vec{B}_0 . \nabla \left( \nabla^2 \tilde{\psi} \right) +
\vec{B}_1. \vec{e}_r  \frac{d J_{0z}}{dr} 
\end{equation}

Symmetry with respect to both the axial and the poloidal coordinates,
means that we need only consider a single mode, 
\begin{equation}
\tilde{\psi} = e^{ikz + im\theta} \tilde{\psi}_m(r)
\end{equation}
where $r$, $\theta$, and $z$ are the usual cylindrical coordinates,
$k$ is a dimensional mode number, and $m$ is a non-dimensional
poloidal mode number. 
Hence we have 
\begin{equation}
\nabla \tilde{\psi} = e^{ikz + im\theta} \left\{ 
\left[ ik\vec{e}_z + \frac{im}{r} \vec{e}_{\theta} \right]
\tilde{\psi}_m(r)  
+ \vec{e}_r \frac{d\tilde{\psi}_m}{dr} 
\right\}
\end{equation}
and because $\vec{B}_1 = \vec{e}_z \wedge \nabla \tilde{\psi}$, then 
$\vec{B}_1 . \nabla r =-\vec{e}_{\theta}.\nabla \tilde{\psi}$, giving 
\begin{equation}
\vec{B}_1 . \vec{e}_r = - e^{ikz + im\theta} \frac{im}{r} \tilde{\psi}_m 
\end{equation}
We also have that
\begin{equation}
\nabla^2 \tilde{\psi} = \left\{ \left(-k^2 -\frac{m^2}{r^2} \right)
\tilde{\psi}_m + \frac{1}{r} \frac{\partial}{\partial r} r
\frac{\partial \tilde{\psi}_m}{\partial r} \right\} e^{ikz + im\theta}
\end{equation}
Hence Eq. \ref{mstab1} gives 
\begin{equation}\label{eleven}
0 = \left( ikB_z + im \frac{B_p}{r} \right)  \left\{ \left(-k^2
-\frac{m^2}{r^2} \right) 
\tilde{\psi}_m + \frac{1}{r} \frac{\partial}{\partial r} r
\frac{\partial \tilde{\psi}_m}{\partial r} \right\} 
+ B_{1r} \frac{dJ_{0\phi}}{dr} 
\end{equation}

$\nabla . \vec{B}=0$ requires the normal component of $\vec{B}$ to be
continuous across the plasma-vacuum surface, that with $\vec{B}_1 =
\vec{e}_z \wedge \nabla \tilde{\psi}$ requires $\tilde{\psi}_m$ to be
continuous across the surface.  
The perturbed field from $d\tilde{\psi}_m/dr$ may be discontinuous however. 
Also $J_{0z}=J_a$ just inside the plasma, but
$J_{0z}=0$ in the vacuum outside the plasma, so if we integrate from an
arbitrarily small distance inside the plasma surface to an
arbitrarily small distance outside the surface, we get
\begin{equation}\label{twelve}
0 = \left( ik B_z + im\frac{B_p}{r} \right)
\left[\left|\frac{d\tilde{\psi}_m}{dr}  \right|\right] - \frac{im}{r} 
\tilde{\psi}_m \left[\left| J_{0\phi}  \right|\right] 
\end{equation}
where $[| f|]$ indicates the {\sl difference} between $f$ evaluated in
the vacuum just outside of the plasma, and $f$ evaluated just inside
the plasma ($[|f|]$ {\sl does not} equal the integral of $f$ from just
inside to just outside the plasma).
Writing $k=-\frac{n}{R}$ and $q=rB_z/RB_p$, then gives 
\begin{equation}\label{cstab}
0 = \left( \frac{m-nq}{m} \right) 
\frac{ \left[\left|r\frac{d\tilde{\psi}_m}{dr}  \right|\right] }{
  \tilde{\psi}_m } + \frac{r J_a}{B_p} 
\end{equation}
as the condition for marginal stability. 

\section{Tokamak Plasmas}

\subsection{Assumptions}

Here we re-examine the assumptions made in going from Eq. \ref{eleven}
to Eq. \ref{twelve}. To obtain Eq. \ref{twelve}, we allowed the
perturbed fields to be discontinuous, but required the equilibrium
fields to be continuous across the plasma-vacuum boundary. 
Because a discontinuous magnetic field requires a skin
current\cite{Jackson}, we will allow perturbed skin currents, but the
continuous equilibrium fields imply zero equilibrium skin
currents. 
Therefore we assume that:  
(a) there are no equilibrium skin currents, but $\vec{J}_0$ can be
  discontinuous at the plasma-vacuum interface, and (b) perturbations
  to the magnetic field induce surface skin currents. 
The first of these assumptions means that with the exception of
$\vec{J}_0$, the equilibrium quantities  will be 
continuous across the plasma-vacuum boundary.  
So if we integrate along the unit normal to the plasma surface, from a
distance $\epsilon$ just inside the surface to a distance $\epsilon$
just outside the plasma, then we will get  
\begin{equation}
\int^{l+\epsilon}_{l-\epsilon} f(\eta(l')) dl' =
\epsilon f(\eta(l)) \rightarrow 0 \mbox{ as } \epsilon \rightarrow 0 
\end{equation}
where $\eta(l')$ is a path parameterised by $l'$, that is parallel to
the unit normal to the plasma, and where $f$ is a continuous
equlibrium quantity.  
We also get
\begin{equation}
\int_{l-\epsilon}^{l+\epsilon} \frac{df}{dl'} dl' =
f(l+\epsilon)-f(l-\epsilon) \rightarrow 0 \mbox{ as } \epsilon
\rightarrow 0  
\end{equation}
However, because the equlibrium current is discontinuous at the plasma
surface being non-zero within the surface but zero in the vacuum,  
its derivatives act like delta functions, and hence 
\begin{equation}
\int_{l-\epsilon}^{l+\epsilon} 
\frac{d\vec{J}_0}{dl'} dl' = 
\vec{J}_0(l+\epsilon) - \vec{J}_0(1-\epsilon) 
\equiv \left[ \left| \vec{J}_0 \right| \right] 
\end{equation}
and similarly 
\begin{equation}
\int_{l-\epsilon}^{l+\epsilon} f \frac{d\vec{J}_0}{dl'} dl' = 
f(l) \left[ \left| \vec{J}_0 \right| \right] 
\end{equation}
The second of these assumptions, that the perturbation will produce
skin currents at the plasma surface (i.e. that the perturbed magnetic
field can be discontinuous at the perturbed  plasma surface), means
that for a perturbed current $\vec{J}_1$ 
\begin{equation}
\int_{l-\epsilon}^{l+\epsilon} \vec{J}_1(l') dl' \equiv 
\frac{\vec{\sigma}}{RB_p} \neq 0 
\end{equation}
i.e. the skin current acts like a delta function at the plasma
surface. 
The reason for including the factor of $1/RB_p$ will be clear later. 
Similarly if $f$ is a continuous function at the plasma surface, 
\begin{equation}
\int_{l-\epsilon}^{l+\epsilon} f \vec{J}_1(l') dl' = f(l) 
\frac{\vec{\sigma}}{RB_p} 
\end{equation}

A few other remarks are worthwhile. 
Firstly the terms in Eq. \ref{linearised} are evaluated at the
equilibrium surface positions, but give the value of the perturbed
force balance equation {\it at the perturbed  surface}.  
Secondly, the perturbed unit normal $\vec{n} = \vec{n}_0 + \vec{n}_1$,
with $\vec{n}_1 \sim |\xi| \ll 1$. Hence we have that  
\begin{equation}
\begin{array}{l}
\int_{l-\epsilon}^{l+\epsilon}  f \frac{d\vec{J}_0}{d\psi} dl' 
\\
= \int_{l-\epsilon}^{l+\epsilon} f \frac{ \nabla \psi . \nabla
  \vec{J}_0}{R^2B_p^2} dl'  
\\ 
= \int_{l-\epsilon}^{l+\epsilon} \frac{f}{RB_p} \left( \vec{n} .\nabla
\vec{J}_0 - \vec{n}_1. \nabla \vec{J}_0 \right) dl'  
\\
= \int_{l-\epsilon}^{l+\epsilon} \frac{f}{RB_p} \vec{n} .\nabla
\vec{J}_0 + O (\xi ) 
\\
= \int_{l-\epsilon}^{l+\epsilon} \frac{f}{RB_p} \frac{d\vec{J}_0}{dl'} dl' 
\\
= \frac{f}{RB_p} \left[ \left| \vec{J}_0 \right| \right] 
\end{array}
\end{equation}
or equivalently, 
\begin{equation}
\int_{l-\epsilon}^{1+\epsilon} f \frac{d\vec{J}_0}{d\psi} dl'
= \frac{1}{RB_p} 
\int_{\psi_{-}}^{\psi_{+}} f \frac{d\vec{J}_0}{d\psi} d\psi + O(\xi) 
\end{equation} 
where $\int_{\psi_{-}}^{\psi_{+}}$ indicates an integral from just inside
the last closed flux surface, to just outside it. 
Similarly because $\nabla \psi .\vec{n} = \nabla \psi .\vec{n}_0 +
O(\xi) = RB_p + O(\xi)$, then 
\begin{equation}
\int_{l-\epsilon}^{1+\epsilon} f dl = \frac{1}{RB_p}
\int_{\psi_{-}}^{\psi_{+}} f
d\psi + O(\xi) 
\end{equation}
and 
\begin{equation}
\vec{\sigma} = \int_{\psi_-}^{\psi_+} J_1 d\psi  + O (\xi^2)
\end{equation}

Bearing the above remarks in mind, we now integrate
Eq. \ref{linearised} across the plasma surface,
distinguishing perturbed quantities by a subscript of $1$ and the 
equilibrium currents and magnetic field by a subscript of $0$, 
to get  
\begin{equation}
0 = \vec{B}_0 . \nabla 
\int^{\psi_+}_{\psi_-} \vec{J}_1 d\psi +
\int^{\psi_+}_{\psi_-} \left( \vec{B}_1
. \nabla \psi \right) \frac{\partial \vec{J}_0}{\partial \psi} d\psi
- \left( \int^{\psi_+}_{\psi_-} \vec{J}_1 d\psi \right) . \nabla
\vec{B}_0  + O\left( \xi^2 \right)
\end{equation}
Note that because the terms in Eq. \ref{linearised} are of order
$\xi$ (it describes a linearised perturbation to the plasma), the
$O(\xi)$ corrections that arise when integrating along the normal to
the surface produce terms of order $\xi^2$ and are neglected. 
Therefore at leading order we have  
\begin{equation}\label{PMeqn}
0= \vec{B}_0 . \nabla \vec{\sigma}_1 + B_1^{\psi} \left[\left|
  \vec{J}_0 \right|\right] - \vec{\sigma}_1 . \nabla \vec{B}_0 
\end{equation}
where $B_1^{\psi}=\vec{B}_1 . \nabla \psi$. This is the generalised
force balance equation for Peeling mode marginal stability, valid for
an arbitrary cross-section Tokamak plasma. The last term is a new term
that is not present in a circular cross-section cylindrical geometry.

The procedure of integrating across the plasma-vacuum boundary is
clearer if we project out the components before integrating across the
surface. 
We have done this as a check, but it is algebraically cumbersome, and
obscures the physical arguments that are clearer in the presentation
above. 


\subsection{Ampere's law at the surface} 

Before proceeding it is worth examining some relationships between
the magnetic fields and the skin currents. 
At the plasma-vacuum interface, $\nabla . \vec{B} =0$ and Ampere's law
imply that\cite{Jackson}
\begin{equation}\label{nB}
\left[\left| \vec{n}.\vec{B} \right|\right]=0
\end{equation}
\begin{equation}
\vec{n} \wedge \left( \vec{B}_V - \vec{B} \right)= \frac{\vec{\sigma}}{RB_p} 
\end{equation}
with $\vec{\sigma}=\int_{\psi_-}^{\psi_+} \vec{J}_1 d\psi$
as before,  
$\vec{n}$ denotes the unit normal to the surface, and
$\vec{B}_V$ the magnetic field in the vacuum. 
Because we assume zero equilibrium skin currents 
\begin{equation}\label{s0}
\frac{\vec{\sigma}_0}{RB_p} = \vec{n}_0 \wedge
  \left( \vec{B}_0^V - \vec{B}_0 \right) = 0 
\end{equation}
and therefore because Eq. \ref{nB} gives $\vec{n}_0 . \left( \vec{B}^V_0
  - \vec{B}_0 \right) = 0$, it then follows that $\vec{B}^V_0 - \vec{B}_0$
  is not parallel to $\vec{n}_0$, and hence the only way to satisfy
  Eq. \ref{s0} is if $\vec{B}_0^V=\vec{B}_0$.  

Considering the lowest order perturbation, we have 
\begin{equation}
\begin{array}{ll}
\frac{\vec{\sigma}}{RB_p} &= \vec{n}_1 \wedge \left(\vec{B}^V_0 -
\vec{B}_0 \right) + \vec{n}_0 \wedge \left( \vec{B}^V_1 - \vec{B}_1
\right) 
\\
&+ \xi. \nabla \left( \vec{n}_0 \wedge \left( \vec{B}^V_0 - \vec{B}_0 
\right) \right) 
\end{array}
\end{equation}
which using $\vec{B}_0^V=\vec{B}_0$ and $\nabla \psi = RB_p\vec{n}_0$,
gives 
\begin{equation}
\vec{\sigma} = \nabla \psi \wedge \left( \vec{B}^V_1 - \vec{B}_1
\right) 
\end{equation}
and hence 
\begin{equation}
\nabla \psi . \vec{\sigma} = 0 
\end{equation}

In the $\psi$, $\chi$, $\phi$ coordinate system it is possible to
directly evaluate $\vec{\sigma}=\int_{l-\epsilon}^{l+\epsilon}  \nabla \wedge
\vec{B}_1 dl'$, giving 
\begin{equation}\label{sc1}
\vec{\sigma} = R^2 \nabla \phi [| \vec{B}_p . \vec{B}_1 |] - R^2
  \vec{B}_p [| \nabla 
  \phi . \vec{B}_1 |] 
\end{equation}

\subsection{Skin currents at marginal stability}

Using $\nabla \psi . \vec{\sigma} =0$, while projecting out
components of Eq. \ref{PMeqn}, gives 
\begin{equation}\label{23}
\begin{array}{ll}
0&= \vec{B}_0. \nabla \left( \vec{\sigma}. \vec{B}_p \right) + 
\vec{B}_1.\nabla{\psi} \left[ \left| \vec{B}_p . \vec{J}_0 \right|\right] -
\vec{\sigma} . \nabla B_p^2 
\\
0&= \vec{B}_0. \nabla \left( \vec{\sigma}. \nabla \phi \right) + 
\vec{B}_1.\nabla {\psi} \left[ \left| \nabla \phi . \vec{J}_0 \right|\right]
+ 2I \frac{\vec{\sigma} .\nabla R}{R^3} 
\end{array}
\end{equation}
Because $2I \frac{\sigma.\nabla R}{R^3} = - \sigma . \nabla
\left(\frac{I}{R^2} \right) = - \left( \frac{\sigma. \vec{B}_p}{B_p^2}
\right) \vec{B}_p . \nabla \left( \frac{I}{R^2} \right)$, then we can
rewrite Eq. \ref{23} as 
\begin{equation}\label{eqs1}
\begin{array}{ll}
0&= \vec{B}_0. \nabla \left( \frac{\vec{\sigma}. \vec{B}_p}{B_p^2} \right) + 
\vec{B}_1.\nabla {\psi} \frac{ \left[ \left|  \vec{B}_p. \vec{J}_0
    \right|\right]}{B_p^2} 
\\
0&= \vec{B}_0. \nabla \left( \vec{\sigma}. \nabla \phi \right) + 
\vec{B}_1. \nabla \psi \left[ \left|  \nabla \phi .\vec{J}_0 \right|\right] 
- \left( \frac{\sigma. \vec{B}_p}{B_p^2} \right) \vec{B}_p. \nabla \left(
\frac{I}{R^2} \right) 
\end{array}
\end{equation}

Next we note that for $\vec{B}_0 = I(\psi) \nabla \phi + \nabla \phi
\wedge \nabla \psi$, then 
\begin{equation}\label{JpJphi} 
\begin{array}{l}
\frac{\vec{B}_p.\vec{J}_0}{B_p^2} = - I'
\\
\nabla \phi . \vec{J}_0 = - p' -\frac{II'}{R^2} 
\end{array}
\end{equation}
giving   
\begin{equation}
\begin{array}{l}
\left[\left| \frac{\vec{B}_p.\vec{J}_0}{B_p^2} \right|\right] =  I_a'
\\
\left[\left| \nabla \phi . \vec{J}_0 \right|\right] =  p_a' +
\frac{I_aI_a'}{R^2}  
\end{array}
\end{equation}
Using $\vec{B}_1 . \nabla \psi = \vec{B}_0 . \nabla \xi_{\psi}$ and
also noting that $I=I(\psi)$, Eqs. \ref{eqs1} now become
\begin{equation}\label{eqs2}
\begin{array}{ll}
0&= \vec{B}_0. \nabla \left( \frac{\vec{\sigma}. \vec{B}_p}{B_p^2} +
I_a'\xi_{\psi} \right) 
\\
0&= \vec{B}_0. \nabla \left( \vec{\sigma}. \nabla \phi 
+ \xi_{\psi} \left( p_a' + \frac{I_aI_a'}{R^2} \right) \right) 
- \vec{B}_0 .\nabla \left( \frac{I}{R^2} \right) 
\left( \frac{\vec{\sigma}. \vec{B}_p}{B_p^2} + I_a'\xi_{\psi} \right)
\end{array}
\end{equation}
Hence we have that
\begin{equation}\label{eq3}
\begin{array}{l}
\frac{\vec{\sigma}.\vec{B}_p}{B_p^2} = - I_a' \xi_{\psi} +
f\left(\phi-\int^{\chi}\nu d\chi' \right) 
\\
\vec{\sigma} . \nabla \phi = - \left(p_a' + \frac{I_aI_a'}{R^2}
\right)\xi_{\psi} + \frac{I}{R^2} f\left(\phi-\int^{\chi}\nu d\chi' 
\right)  
\end{array}
\end{equation}
where $f$ is some function of $\phi - \int^{\chi} \nu d\chi'$, so that
$\vec{B}_0 . \nabla f =0$. Therefore using Eq. \ref{JpJphi} to write
Eqs. \ref{eq3} in terms of the equilibrium current $\vec{J}_0$, we
have the skin currents at marginal stability given by  
\begin{equation}\label{mstabsigma}
\vec{\sigma} = \vec{J}_0 \xi_{\psi} + \vec{B}_0 f\left(\phi-\int^{\chi}\nu
d\chi' \right) 
\end{equation}

If we require that $\vec{\sigma}=0$ for $\xi_{\psi}=0$, then because
$\vec{B}_0 .\nabla \xi_{\psi} \neq 0$, we simply have that 
\begin{equation}\label{sigma}
\vec{\sigma} = \vec{J}_0 \xi_{\psi} 
\end{equation}
i.e. that the skin current due to a perturbation at marginal stability
is equal  to the product of the equilibrium current at the edge and
the radial  displacement of the plasma. 

Notice that at this point we have not used any expansion, e.g. in
terms of straight field line co-ordinates, and we have not neglected
any poloidal dependencies of the equilibrium quantities. We have
implicitly assumed the existence of the $\chi$, $\phi$, $\psi$
coordinate system, but it is possible to re-express the coordinates in
terms of arc length along a flux surface, and the consequent relations
expressed in Eqs. \ref{sc1} for example, remain valid (except
at the point of zero size that is the X-point). Hence the results
appear valid at the separatrix.

\subsection{Marginal stability}

To relate this to the previous cylindrical condition for marginal
stability, Eq. \ref{cstab}, we consider $\nabla . \vec{B}$, which may
be written  
\begin{equation}
\nabla . \vec{B} = \frac{1}{J_{\chi}} \left( \frac{\partial}{\partial
  \psi} \left(J_{\chi}\nabla \psi . \vec{B} \right) 
+ \frac{\partial}{\partial \chi} \left( J_{\chi}\nabla \chi . \vec{B}
  \right) 
+ \frac{\partial}{\partial \phi} \left( J_{\chi} \nabla\phi.\vec{B}
  \right) \right) 
\end{equation} 
with $J_{\chi}$ the Jacobian. 
Then we evaluate $[| \nabla . \vec{B} |]$, the difference between
$\nabla . \vec{B}$ evaluated in the vacuum just outside the plasma and
$\nabla . \vec{B}$ evaluated just inside the plasma. 

Noting that $J_{\chi} \nabla  \chi
.\vec{B}_1=J_{\chi}\frac{1}{J_{\chi}B_p} \frac{\nabla \chi}{|\nabla
  \chi|} . \vec{B}_1 = \frac{\vec{B}_p.\vec{B}_1}{B_p^2}$, then 
$[| \nabla . \vec{B}_1 =0|]$ requires that 
\begin{equation}
0= \left[ \left| \frac{1}{J_{\chi}} \frac{\partial}{\partial \psi}
  J_{\chi} \nabla \psi . \vec{B}_1 \right| \right]
+\frac{1}{J_{\chi}} \frac{\partial}{\partial \chi} \left[ \left|
  \frac{\vec{B}_p . \vec{B}_1}{B_p^2} \right| \right] 
+ \frac{\partial}{\partial \phi} \left[\left| \nabla \phi . \vec{B}_1
  \right| \right] 
\end{equation}
Eq. \ref{sc1} gives 
\begin{equation}
\begin{array}{l}
-\frac{\vec{\sigma}.\vec{B}_p}{R^2B_p^2} = \left[ \left| \nabla\phi
 . \vec{B}_1 \right| \right] 
\\
\vec{\sigma} . \nabla \phi = \left[ \left| \vec{B}_p . \vec{B}_1
 \right| \right] 
\end{array}
\end{equation}
So we have
\begin{equation}
0 = \left[ \left| \frac{1}{J_{\chi}} \frac{\partial}{\partial \psi}
  J_{\chi} \nabla \psi . \vec{B}_1 \right| \right] 
+ \frac{1}{J_{\chi}} \frac{\partial}{\partial \chi} \left(
  \frac{\vec{\sigma} . \nabla \phi}{B_p^2} \right) 
+ \frac{\partial}{\partial \phi} \left(
  -\frac{\vec{\sigma}.\vec{B}_p}{R^2B_p^2} \right) 
\end{equation}

To simplify this we consider a limit of high toroidal mode number $n$ and
note that $\frac{1}{J_{\chi}}\frac{\partial}{\partial \chi} =
\vec{B}_0 
. \nabla - \frac{I}{R^2} \frac{\partial}{\partial \phi}$. Then we note
that 
$\vec{B}_0 . \nabla$ is of order $1$ to prevent a large stabilising
contribution from field-line bending, but $\frac{\partial}{\partial
  \phi}$ is of order $n$. 
We also have $\frac{\partial}{\partial \psi}
\nabla \psi . \vec{B}_1 \sim n$.
Because the system is axisymmetric, we need only consider a single
Fourier mode in the toroidal angle, and 
take $\vec{\sigma} \sim e^{-in\phi}$. 
Then we have 
\begin{equation}
0 = \left[ \left| \frac{1}{J_{\chi}} \frac{\partial}{\partial \psi}
  J_{\chi} \nabla \psi . \vec{B}_1 \right| \right] 
+ \frac{in}{R^2 B_p^2} \left( I \vec{\sigma}. \nabla \phi +
  \vec{\sigma} . \vec{B}_p \right) + O(1) 
\end{equation}
Later in Section \ref{apnb} we will find that $[| \nabla
  \psi. \vec{B}_1|]=0$,  so there will be zero contribution from the
  terms  involving $\partial J_{\chi}/\partial \psi$. Rearranging the
  resulting equation leaves 
\begin{equation}\label{feq1}
R^2B_p^2 \left[ \left|  \frac{i}{n} 
\frac{\partial}{\partial \psi}  
  \nabla \psi . \vec{B}_1 \right| \right] 
= \vec{B}_0 . \vec{\sigma} + O \left( \frac{1}{n} \right) 
\end{equation}
Eq. \ref{feq1} is our generalised criterion for marginal stability to
the Peeling mode at high-$n$, with $\vec{\sigma}$ given by
Eq. \ref{sigma}.

\subsection{Cylindrical limit}

Using Eq. \ref{sigma} ($\vec{\sigma}=\xi_{\psi} \vec{J}_0=
\vec{J}_0 \nabla \psi . \vec{\xi}$), then for  
large aspect ratio (cylindrical) geometry $\vec{\sigma} = \vec{J}_0
(RB_p)  \xi_r$ and 
Eq. \ref{feq1} becomes 
\begin{equation}
\frac{RB_p}{rB_0} \left[ \left| \frac{i}{n} \frac{d b_r}{dr}  \right| \right]
= \frac{\vec{B}_0 . \vec{J}_0}{B_0} \frac{\xi_r}{r} 
\end{equation}
Noting that $b_r = \vec{B}.\nabla \xi_r$, taking $\xi
\sim e^{im\theta - in\phi}$, and using $\frac{1}{nq}=\frac{1}{m} +
\frac{m-nq}{nq}$ with $m\simeq nq$, this gives 
\begin{equation}
0= \frac{r J_{\parallel}}{B_p} + \frac{m-nq}{m} \frac{\left[ \left| r
    \frac{db_r}{dr} \right| \right]}{b_r} 
\end{equation}
where we have also used that for a cylinder $q=\frac{rB_0}{R_0B_p}$. 
Finally the plasma vacuum boundary conditions (e.g. see
Freidberg\cite{Freidberg}),  of  
\begin{equation}\label{BCs1}
\vec{n}_0.\vec{B}_1^V = \vec{B}_0 . \nabla \left( \vec{n}_0.\vec{\xi}
\right)  - \left( \vec{n}_0 . \vec{\xi} \right) 
\vec{n}_0 . \vec{n}_0 . \nabla \vec{B}_0 
\end{equation}
for a cylinder simplify to give 
$b_r^V = \vec{B}_0 .\nabla \xi_r = b_r$. So we regain the
usual condition for marginal stability to the Peeling mode in
cylindrical geometry, of 
\begin{equation}\label{dp1}
\Delta'_a \Delta_a + J_{\parallel} = 0 
\end{equation}
with $\Delta'_a = \left[ \left| \frac{r}{b_r} \frac{db_r}{dr} \right|
  \right]$ and $\Delta_a = \left( 1 - \frac{nq}{m} \right)$.

\section{The Plasma-Vacuum Boundary Conditions}\label{apnb}

The question now arises: what are the equivalent terms in the energy
principle that correspond to our ordering for the Peeling mode?
Before addressing this question, we first show that the plasma-vacuum 
boundary condition that is usually given in conjunction with the
energy principle simply requires continuity of the normal component
of the perturbed magnetic field, evaluated with the equilibrium normal
to the surface at the equilibrium surface position. 
This is shown to agree with a simpler and more intuitive derivation
that is given later. 

We start from the boundary condition usually given in conjunction with
the energy principle (Freidberg\cite{Freidberg}), of 
\begin{equation}
\vec{n}_0 . \vec{B}_1^V = 
\vec{B}_0 . \nabla \left( \vec{n}_0 . \vec{\xi} \right) 
- \left( \vec{n}_0 . \vec{\xi} \right) 
\left(
\vec{n}_0 .\left( \vec{n}_0 . \nabla \right) \vec{B}_0  \right)
\end{equation}
with $\vec{n}_0$ the equilibrium unit normal. We will rewrite the
expression in terms of $\nabla \psi = \vec{n}_0 RB_p$, 
and then simplify the result. Firstly we
multiply by $RB_p$, and rearrange the equation to get 
\begin{equation}
\begin{array}{ll}
\nabla \psi . \vec{B}_1^V 
&= RB_p \vec{B}_0 . \nabla \left( \vec{n}_0 . \vec{\xi} \right) 
- \left( \nabla \psi . \vec{\xi} \right) 
\left( \frac{1}{R^2B_p^2} 
\nabla \psi . \nabla \psi . \nabla \vec{B}_0  
\right)
\\
&=  \vec{B}_0 . \nabla \left( RB_p \vec{n}_0 . \vec{\xi} \right) 
- \left( \vec{n}_0 . \vec{\xi} \right) \vec{B}_0 . \nabla \left( RB_p
\right) 
\\
&- \frac{\nabla \psi . \vec{\xi}}{R^2 B_p^2} 
\left[ 
\nabla \psi . \nabla \left( \vec{B}_0 . \nabla \psi \right) 
- \vec{B}_0 . \nabla \psi . \nabla \psi \nabla \psi \right]
\\
&=  \vec{B}_0 . \nabla \left( \nabla \psi . \vec{\xi} \right) 
- \frac{\nabla \psi . \vec{\xi}}{RB_p} \vec{B}_0 . \nabla \left( RB_p
\right) 
- \frac{\nabla \psi . \vec{\xi}}{R^2 B_p^2} 
\left[ - \nabla \psi . \vec{B}_0 . \nabla \psi \nabla \psi \right]
\\
&=  \vec{B}_0 . \nabla \left( \nabla \psi . \vec{\xi} \right) 
- \frac{\nabla \psi . \vec{\xi}}{RB_p} \vec{B}_0 . \nabla \left( RB_p
\right) 
- \frac{\nabla \psi . \vec{\xi}}{R^2 B_p^2} 
\left[ - \vec{B}_0 . \nabla \left( \frac{R^2B_p^2}{2} \right) \right] 
\\
&=  \vec{B}_0 . \nabla \left( \nabla \psi . \vec{\xi} \right) 
\\
&=  \nabla \psi . \vec{B}_1 
\end{array}
\end{equation}

Alternately, $\nabla . \vec{B} =0$ at the surface requires
$\vec{n}. \vec{B} = \vec{n}. \vec{B}_V$, with $\vec{n}=\vec{n}_0 +
\vec{n}_1$, where $\vec{n}_0$ is the unit normal to the equilibrium
magnetic  flux surfaces (and hence also the equilibrium plasma
surface) and  $\vec{n}_1$ is the unit normal perturbation from the
equilibrium  unit normal $\vec{n}_0$ (and hence also the perturbation
from  the equilibrium unit normal to the plasma surface). 
Then $\vec{n}. \vec{B} = \vec{n}. \vec{B}_V$ requires 
\begin{equation}\label{neq1}
\left( \vec{n}_0 + \vec{n}_1 \right) . \left( \vec{B}_0 + \vec{B}_1
\right) + \vec{\xi} . \nabla \left( \vec{n}_0 . \vec{B}_0 \right) =
\left( \vec{n}_0 + \vec{n}_1 \right) . \left( \vec{B}_0^V +
\vec{B}_1^V \right) + \vec{\xi} . \nabla \left( \vec{n}_0
. \vec{B}_0^V \right) 
\end{equation}
where $\vec{\xi}$ is the displacement of the plasma from its
equilibrium  position, and where all quantities are evaluated at their
equilibrium positions. 
Similarly at equilibrium it is required that  
\begin{equation}
\vec{n}_0 . \vec{B}_0 = \vec{n}_0 . \vec{B}_0^V 
\end{equation}
again with the quantities evaluated at their equilibrium positions. 
Therefore because $\vec{n}_0.\vec{B}_0 = \vec{n}_0 .\vec{B}_0^V=0$
everywhere,  then $\vec{\xi} .\nabla \left( \vec{n}_0 . \vec{B}_0
\right) =  \vec{\xi} . \nabla \left( \vec{n}_0 . \vec{B}_0^V \right) =
0$,  and Eq. \ref{neq1} may be simplified and rearranged as 
\begin{equation}\label{neq1}
\vec{n}_1 . \left( \vec{B}_0 - \vec{B}_0^V \right) 
+ \vec{n}_0 . \left( \vec{B}_1 - \vec{B}_1^V \right) 
 = 0
\end{equation}
Assuming there are no skin currents at equilibrium, then as shown
previously in the main text, we have $\vec{B}_0=\vec{B}_0^V$, so
Eq. \ref{neq1} becomes 
\begin{equation}
\vec{n}_0 . \left( \vec{B}_1 - \vec{B}_1^V \right) = 0
\end{equation}
or upon multiplying both sides by $RB_p$ 
\begin{equation}
\nabla \psi . \left( \vec{B}_1 - \vec{B}_1^V \right) = 0
\end{equation}
as above.

\section{The Energy Principle}\label{EnergyP1}

Now we return to the relationship between our equations for marginal
stability of the Peeling mode, and the energy principle. 
We start from the high mode number formulation for $\delta W= \delta
W_F + \delta W_S + \delta W_V$, given in Connor et
al\cite{Connor}. Looking 
firstly at $\delta W_S$, then later $\delta W_V$, we
have
\begin{equation}
\delta W_S = \pi \oint d\chi \frac{\xi^*}{n} J_{\chi}B k_{\parallel} 
\left[ 
\frac{R^2B_p^2}{J_{\chi}B^2} \frac{1}{n} \frac{\partial}{\partial \psi}
\left( J_{\chi}B k_{\parallel} \xi_{\psi} \right) 
-\frac{\vec{B}.\vec{J}}{B^2} \xi_{\psi} 
\right]
\end{equation} 
with $J_{\chi}B k_{\parallel} = iJ_{\chi} \vec{B}.\nabla$. 
Note that Connor et al\cite{Connor} took $\xi \sim e^{+in\phi}$,
however here we have $\xi \sim e^{-in\phi}$, which is reflected by 
$n\rightarrow -n$ in the expression for $\delta W$.
We integrate by parts, 
noting that $\xi \sim e^{-in\phi}$ but $\xi^* \sim e^{in\phi}$, 
and also replace $J_{\chi}B k_{\parallel}$ with
$iJ_{\chi} \vec{B}.\nabla$ to give 
\begin{equation}
\delta W_S = -\pi \oint J_{\chi} d\chi \left( \frac{i}{n}
\vec{B}.\nabla \xi^* \right) 
\left[ 
\frac{R^2B_p^2}{J_{\chi}B^2} \frac{i}{n} \frac{\partial}{\partial \psi}
\left( J_{\chi} \vec{B}. \nabla \xi_{\psi} \right) 
-\frac{\vec{B}.\vec{J}}{B^2} \xi_{\psi} 
\right]
\end{equation}
Using $\vec{B}.\nabla \xi_{\psi}^* = \nabla \psi . \vec{B}_1^*$,
then we get 
\begin{equation}
\delta W_S = \pi \oint J_{\chi} d\chi \left(\frac{i}{n} \nabla \psi
. \vec{B}_1^* \right) 
\left[ 
- \frac{R^2B_p^2}{B^2} \frac{i}{n} \frac{1}{J_{\chi}} 
\frac{\partial}{\partial \psi} 
\left( J_{\chi} \nabla\psi . \vec{B}_1 \right) 
+\frac{\vec{B}.\vec{J}}{B^2} \xi_{\psi} 
\right]
\end{equation}

To obtain the vacuum solution for $\vec{B}_1^V = \nabla V$, we need to
solve $\nabla^2 V =0$, which requires 
\begin{equation}
0= \frac{1}{J_{\chi}} \left\{ \frac{\partial}{\partial \psi} \left(
J_{\chi} \nabla \psi . \nabla V \right) 
+ \frac{\partial}{\partial \chi} \left( J_{\chi} \nabla \chi
. \nabla V \right)  
+ \frac{\partial}{\partial \phi} \left( J_{\chi} \nabla \phi 
. \nabla V \right)  \right\}
\end{equation}
Using $\frac{1}{J_{\chi}}\frac{\partial}{\partial \chi} = \left(
  \vec{B} . \nabla - \frac{I}{R^2} \frac{\partial}{\partial \phi}
  \right)$, $\left| \nabla \chi \right|^2 = \frac{1}{J_{\chi}^2B_p^2}$,
  and $V \sim e^{-in\phi}$, we can rewrite this as 
\begin{equation}
0= \frac{1}{J_{\chi}} \frac{\partial}{\partial \psi} 
\left( J_{\chi} \nabla \psi . \nabla V \right) + \left( \vec{B}
. \nabla + in\frac{I}{R^2} \right) \frac{1}{B_p^2} 
 \left( \vec{B}. \nabla + in\frac{I}{R^2} \right)V
- \frac{n^2}{R^2}
\end{equation}
Then using $\vec{B}. \nabla \sim 1$, and taking the high-$n$ limit, gives 
\begin{equation}
0 = \frac{1}{J_{\chi}} \frac{\partial}{\partial \psi} \left( J_{\chi} 
\nabla \psi . \nabla V \right) 
- n^2 \frac{I^2}{R^4B_p^2} V - n^2 \frac{1}{R^2} V 
\end{equation}
which using $\vec{B}_1^V=\nabla V$, rearranges to give the result that
for $n\gg 1$,  
\begin{equation}\label{EqV} 
V = \frac{R^2B_p^2}{B^2} \frac{1}{n^2} \frac{1}{J_{\chi}} 
\frac{\partial}{\partial \psi}
\left( J_{\chi} \nabla \psi . \vec{B}_1^V \right)
\end{equation}

The vacuum contribution to $\delta W$ is 
\begin{equation}
\delta W_V = \frac{1}{2} \int \vec{dr} \left| \vec{B}_1^V \right|^2 
\end{equation}
 Using $\nabla^2 V = 0$, $\vec{B}_1 = \nabla V$, and Gauss' theorem,
this may be written as an integral over the plasma surface, with 
\begin{equation}
\delta W_V = - \frac{1}{2} \int V \vec{dS} .  \nabla V^* 
\end{equation}
Using $\vec{dS} = \left( \frac{\nabla \psi}{RB_p} \right) \left(
J_{\chi} B_p d\chi \right) \left( Rd\phi \right)$, and integrating
with respect to $\phi$ gives 
\begin{equation}
\delta W_V = - \pi \oint J_{\chi} d\chi V \nabla \psi . \nabla V^* 
\end{equation}
Using the expression Eq. \ref{EqV} for $V$ at high-$n$, and the
boundary condition $\nabla \psi . \vec{B}_1^V = \nabla \psi
. \vec{B}_1$ (that Section \ref{apnb} shows to be equivalent to the 
boundary condition that is usually given in
formulations of the energy principle), we get 
\begin{equation}
\delta W_V = - \pi \oint J_{\chi} d\chi \left( \nabla \psi . \vec{B}_1
\right) \left[  \frac{R^2B_p^2}{B^2} \frac{1}{n^2} \frac{1}{J_{\chi}} 
  \frac{\partial}{\partial \psi} 
\left( J_{\chi} \nabla \psi . \vec{B}_1^V \right) \right]
\end{equation}
Inserting $-1=i^2$ into $\delta W_V$, we get 
\begin{equation}
\begin{array}{ll}
\delta W_S + \delta W_V = 
\pi \oint J_{\chi} d\chi \left( \frac{i}{n} \right) 
\left( \nabla \psi . \vec{B}_1^* \right) 
&\left[ 

\frac{R^2B_p^2}{B^2} \frac{i}{n} \frac{1}{J_{\chi}}
\frac{\partial}{\partial  \psi} 
\left( J_{\chi} \nabla \psi . \vec{B}_1^V \right) \right. 
\\
& \left. 
- \frac{R^2B_p^2}{B^2} \frac{i}{n} \frac{1}{J_{\chi}}
\frac{\partial}{\partial  \psi}
\left( J_{\chi} \nabla\psi . \vec{B}_1 \right) 
+\frac{\vec{B}.\vec{J}}{B^2} \xi_{\psi} 
\right]
\end{array}
\end{equation}
Because $\nabla \psi.\vec{B}_1^V = \nabla \psi . \vec{B}_1$, the terms
involving $\partial J_{\chi}/\partial \psi$ will cancel. 
Then using the notation $[|f|]$ to denote the difference between
$f$ evaluated just outside and just inside the plasma, gives 
\begin{equation}\label{eppm}
 \delta W_S + \delta W_V = 
\pi \oint J_{\chi} d\chi \left( \frac{i}{n} \right) 
\frac{\left( \nabla \psi . \vec{B}_1^* \right)}{B^2} 
\left\{ 
R^2B_p^2 \left[ \left| \frac{i}{n} \frac{\partial}{\partial \psi} 
\left( \nabla \psi . \vec{B}_1 \right) \right| \right]
+\vec{B}.\vec{J} \xi_{\psi} 
\right\}
\end{equation}
The term in \{\} is exactly Eq. \ref{feq1} with $\vec{\sigma} =
\vec{J}_0\xi_{\psi}$ as given by Eq. \ref{sigma}. Therefore 
marginal stability of the Peeling mode corresponds (at high-$n$) to
taking the plasma's contribution to $\delta W$, $\delta W_F\equiv 0$, and
then solving $\delta W_S + \delta W_V = 0$. 
This suggests we should define 
the high-$n$ Peeling mode as a mode (represented by a trial function
in this analysis), that allows us to neglect 
$\delta W_F$ compared to $\delta W_S + \delta W_V$ (by for example
being sufficiently localised), and whose 
subsequent stability is determined by $\delta W_S$ and $\delta W_V$.

\section{X-Point Plasmas}  

Now we consider the stability of Peeling modes to the trial function
considered by Laval et al\cite{Laval}, that consists of a single
Fourier mode with 
$\xi=\xi_m(\psi) e^{im\theta-in\phi}$,  where
$\theta =\frac{1}{q}\int^{\chi}\nu d\chi'$ is the usual straight
field-line poloidal coordinate. 
After taking Eq. \ref{eppm}, and using 
$\nabla\psi .\vec{B}_1^V = \nabla\psi . \vec{B}_1=\vec{B} .\nabla
\xi_{\psi}$, we have 
\begin{equation}\label{e1}
\delta W_S + \delta W_V = \pi \oint J_{\chi} d\chi \left( \frac{i}{n}
\right) \frac{\vec{B}.\nabla \xi_{\psi}^*}{B^2} 
\left\{
R^2B_p^2 \left[ \left| \frac{\frac{\partial}{\partial \psi} \nabla
    \psi .\vec{B}_1}{\nabla \psi \vec{B}_1} \right| \right] 
\left( \frac{i}{n} \right) \vec{B} . \nabla \xi_{\psi} 
+  \vec{B}.\vec{J} \xi_{\psi} 
\right\}
\end{equation}
Substituting the trial function into Eq. \ref{e1} gives  
\begin{equation}\label{W'}
\delta W = -2\pi^2 \frac{\left| \xi_m \right|^2}{R_0} \Delta \left( \Delta
\hat{\Delta}' + \hat{J} \right) 
\end{equation}
where 
\begin{equation}
\Delta \equiv \frac{m-nq}{nq}
\end{equation}
\begin{equation}
\hat{J} \equiv \frac{1}{2\pi} \oint dl \frac{IR_0}{R^2B_p}
\frac{\vec{J}.\vec{B}}{B^2}
\end{equation}
\begin{equation}
\hat{\Delta}' \equiv \left[ \left|
 \frac{1}{2\pi} \oint dl R_0 B_p \frac{I^2}{R^2B^2}
\frac{\frac{\partial}{\partial \psi} 
\left( \nabla \psi    .\vec{B}_1 \right)}{\nabla \psi .\vec{B}_1}
 \right| \right] 
\end{equation}
and with
$dl=J_{\chi}B_pd\chi$ an
element of arc length in the poloidal cross-section, and $R_0$ a
typical  measure of the major radius such as it's average for example. 
Note that because $\xi_m$ is a Fourier component of 
$\vec{\xi}.\nabla \psi \sim \xi (RB_p)$, the dimensions of
$|\xi_m|^2/R_0$ are  energy. 
Equation \ref{W'} may easily be minimised for $\Delta$ (or
equivalently,  minimised with respect to choice of toroidal mode
number), with 
\begin{equation}\label{dmin}
\Delta = -\frac{\hat{J}}{2\hat{\Delta}'}
\end{equation}
giving 
\begin{equation}
\delta W = \left( \frac{\pi}{2} \right)^2 \frac{\left| \xi_m
  \right|^2}{R_0}  \left( 
\frac{2\hat{J}^2}{\hat{\Delta}'} \right)
\end{equation}  
If $\Delta$ is chosen to
maximise the growth rate, then a similar but different value will be
found.   
Similarly, there is  no reason why there
should not be a more unstable mode than the  trial function we have
considered. 
However, our primary interest is stability to the trial function that
was found to be unstable by Laval et al\cite{Laval}. 


A calculation of $\delta W$ requires the evaluation of $\Delta'$ for a
plasma equilibrium with a separatrix. 
The calculation of $\Delta'$ is the main subject of the second part to
this paper. 
As an introduction to this, we note that for a circular cross section
plasma, an estimate for $\Delta'$ may be found by approximating the 
perturbation to the magnetic field near the edge of the plasma as
being the  same as for a  vacuum, then solving  Laplace's equation
both inside  and outside the plasma, and matching the solutions at the
plasma-vacuum  boundary. 
Then for a circular cross-section $\Delta' =-2m \simeq -2nq$.

Observe that for $\vec{J} .\nabla \phi =0$, $\vec{B}_p
. \vec{J}=-I'B_p^2$,  for which 
\begin{equation}
\hat{J} = \frac{1}{2\pi} \oint dl\frac{IR_0}{R^2B_p}
\frac{(-I')B_p^2}{B^2}  \sim 1
\end{equation}   
and as discussed above we will have $\delta W \rightarrow 0$ as $q
\rightarrow  \infty$. 
However, for $\vec{J} . \nabla \phi \neq 0$, then $\vec{B} . \vec{J} =
-Ip' -  B^2I'$ and therefore 
\begin{equation}
\begin{array}{ll}
\hat{J} & = \frac{1}{2\pi} \oint dl \frac{IR_0}{R^2B_p}
\frac{(-Ip'-B^2I')}{B^2}  
\\
& \simeq \left( R_0\frac{\vec{J} . \vec{B}}{B^2} \right) \oint
\frac{I}{R^2} \frac{dl}{B_p}  
\\
& = R_0 \left( \frac{\vec{J} . \vec{B}}{B^2} \right) q
\end{array}
\end{equation} 
For which if $\Delta' \sim -nq$ as is the case for a circular
cross-section, then $\hat{J}^2/\Delta'$ would be of order $-q$, and
$\delta W <0$, suggesting that the mode would be unstable.

Although the sign of $\delta W$ is usually taken to indicate whether a
mode is unstable or not, the growth rate determines how unstable the
mode is (i.e. how  rapidly it develops). 
For example if our trial function $\xi_m(\psi)e^{im\theta} =
\xi_{\psi} =  \nabla \psi . \vec{\xi}$ had been
$\xi_m(\psi)e^{im\theta}  =\nabla \psi .\vec{\xi} / RB_p$ so that it
had  dimensions of length as opposed to dimensions of length times
$RB_p$, then  we would no longer have $\hat{J} \sim q$,  despite our
model only  depending on the poloidal structure of the mode. 
The dependence of $\delta W$ on the normalisation of the
plasma  perturbation does not affect the calculation of the growth
rate however,  for which the consequences of the normalisation of the
plasma  perturbation will cancel. 
The growth rate is discussed later.

Are there any reasons why a computer code might fail to find an
unstable mode? 
One possibility is that the need for 
$m\sim nq$ will require very high poloidal mode numbers as $q
\rightarrow \infty$, and this could potentially prevent a numerical code
from  seeing the instability. 
Also important, is the need to consider the most unstable mode. 
Minimising $\delta W$ with respect to the toroidal mode number gave
$\Delta=-\hat{J}/2\Delta'$. 
If $\Delta' \simeq -2nq$ and $\hat{J} \simeq q R_0
(\vec{J}.\vec{B})/B^2$  (for $\vec{J}.\nabla \phi \neq 0$), this would
require 
\begin{equation}
\Delta \simeq 
-\frac{\hat{J}}{2\Delta'}  = 
R_0  \frac{\vec{J}.\vec{B}}{B^2} \left( \frac{1}{4n} \right) 
\end{equation}
that is independent of $q$ and the poloidal mode number. 
A final possibility is that $\Delta'$ 
might diverge  more rapidly than $\hat{J}$ as we approach the
separatrix. 
The second part to this paper calculates $\Delta'$ analytically,
thereby avoiding the numerical problems
associated with an X-point.  

Another, less obvious reason why an unstable mode might not be found
in computer calculations is that despite $\delta W<0$ 
indicating that it  is energetically favourable for the mode to be
unstable,  the growth rate can still be vanishingly small. 
This possibility is explored next.

\section{The Growth Rate}\label{grsec}

So far we have only considered $\delta W$, because its sign is usually
presumed to be sufficient to indicate whether a mode is stable or not.  
The growth rate $\gamma$ for a mode with $\xi \sim e^{\gamma t}$ is
obtained  from 
$\gamma^2 = -\delta W/\frac{1}{2} \int \rho_0 |\xi |^2 \vec{dr}$,  
where $\frac{1}{2} \int \rho_0 |\xi |^2 \vec{dr}$ is 
the kinetic  energy term.
Next we will estimate the kinetic energy term, so as to estimate the
growth rate.  
The surprising result that we will find is that even if $\delta W <0$,
indicating it is energetically favourable for an instability, the
kinetic  energy term can diverge so strongly that although
it may be  energetically favourable for a mode to be unstable, its
growth rate is vanishingly small. 
An alternative complementary calculation to the one given below, with
the same conclusions, is given in Appendix \ref{Altderiv}.

To estimate $\int \rho_0 |\vec{\xi}|^2 \vec{dr}$, we write 
\begin{equation}\label{xi}
\vec{\xi} = \xi_{\psi}\frac{ \nabla \psi}{R^2B_p^2} + 
\xi_B
\frac{
  \vec{B}}{B^2}  + 
\xi_{\perp} 
\frac{ \vec{B} \wedge \nabla
  \psi}{R^2B_p^2 B^2} 
\end{equation}
for which 
\begin{equation}\label{modxisqd}
|\vec{\xi}|^2 = 
  \frac{\left| \xi_{\psi} \right|^2}{R^2B_p^2}
+ \frac{\left| \xi_B \right|^2}{B^2} 
+ \frac{\left| \xi_{\perp} \right|^2}{R^2B_p^2B^2} 
\end{equation}
It is convenient to write $\vec{\xi}$ in the form of Eq. \ref{xi} 
so that we can use the results from a high-$n$ ordering (e.g. see
Webster and Wilson\cite{Webster02}), that gives  
\begin{equation}\label{xiperpn}
\xi_{\perp} = \frac{i}{n} \nabla  \psi . \nabla \xi_{\psi} =
\frac{i}{n} R^2B_p^2 \frac{\partial \xi_{\psi}}{\partial \psi}  
\end{equation} 
and 
\begin{equation}\label{xiBn}
\vec{B} .\nabla \xi_B - \xi_B \frac{\vec{B}.\nabla B^2}{B^2} = 
\xi_{\psi} \frac{\partial}{\partial \psi} \left( 2p + B^2 \right) 
+ \frac{I\xi_{\perp}}{R^2B_p^2B^2} \vec{B}.\nabla B^2 
\end{equation}


Before continuing further we make some observations on the high-$n$
ordering that for 
$\nabla .\vec{\xi}=0$ usually leads to  
$\xi_{\perp} =\frac{i}{n} R^2B_p^2 \frac{\partial \xi_{\psi}}{\partial
  \psi}$. 
This analysis is used in the derivation of $\delta W$ used in
Section \ref{EnergyP1} onwards, and to derive the equations solved by
ELITE\cite{Wilson02}. 
The ordering implicitly assumes that
$\frac{\partial \xi_{\psi}}{\partial \psi} \gg
\frac{\xi_{\psi}}{J_{\chi}} \frac{\partial J_{\chi}}{\partial \psi}$,
which is the case in the plasma core where
$\frac{1}{J_{\chi}}\frac{\partial J_{\chi}}{\partial \psi} \sim 1$,
because $\frac{\partial \xi_{\psi}}{\partial \psi} \sim n \gg 1$. 
Whereas a sufficiently large $n$ can always be found to ensure
$\frac{1}{n} \frac{\xi_{\psi}}{J_{\chi}} \frac{\partial
  J_{\chi}}{\partial \psi} \ll 1$, and terms of this type will often
be negligible order one contributions anyhow, future calculations would be
improved by including them. 
For example $\xi_{\perp}$ would then become 
$\xi_{\perp} =\frac{i}{n} \frac{R^2B_p^2}{J_{\chi}}
\frac{\partial (J_{\chi}  \xi_{\psi})}{\partial \psi}$.
For the present we will continue to use the ordering employed by
Connor\cite{Connor}, that is also used to derive the equations solved
by ELITE.

Because the trial function that we consider consists of a single
Fourier mode, Eq. \ref{xiBn} may be solved for $\xi_B$, with 
\begin{equation}
\xi_B = \frac{\xi_{\psi} \frac{\partial}{\partial \psi} \left( 2p +B^2
  \right) + \frac{I\vec{B} .\nabla B^2}{B^2} \left( \frac{i}{n}
  \frac{\partial \xi_{\psi}}{\partial \psi} \right)}
{\frac{I}{qR^2} \left( im - inq \right) - \frac{\vec{B}.\nabla
  B^2}{B^2} }
\end{equation}
giving  
\begin{equation}\label{xiBsqd}
\frac{\left| \xi_B \right|^2}{B^2} = \frac{1}{B^2} 
\frac{ \left| \xi_{\psi} \right|^2 \left( \frac{\partial}{\partial
    \psi} \left( 2p + B^2 \right) \right)^2 + \frac{1}{n^2} 
\left| \frac{\partial \xi_{\psi}}{\partial \psi} \right|^2 I^2 
\left( \frac{\vec{B}.\nabla B^2}{B^2} \right)^2 }
{ \frac{I^2}{R^4} n^2 \Delta^2 + \left( \vec{B .\nabla B^2}{B^2}
  \right)^2} 
\end{equation}
with $\Delta =(m-nq)/nq$ as before. 
Using Eq. \ref{xiperpn} for $\xi_{\perp}$, and substituting this and
\ref{xiBsqd} into Eq. \ref{modxisqd}, gives 
\begin{equation}
\begin{array}{ll}
| \vec{\xi} |^2 
&= \left| \xi_{\psi} \right|^2 
\left\{ \frac{B^2}{R^2B_p^2} 
+ \left. \left( \frac{\partial}{\partial \psi} \left( 2p + B^2 \right)
\right)^2 \right/ \left(\frac{I^2}{R^4}n^2 \Delta^2 + 
\left( \frac{\vec{B}.\nabla B^2}{B^2} \right)^2 \right) \right\} 
\\
&+ \frac{1}{n^2} \left| \frac{\partial \xi_{\psi}}{\partial \psi}
\right|^2 
\left. 
\left\{ R^2 \left( \frac{\vec{B}.\nabla B^2}{B^2} \right)^2 
+ \frac{I^2B_p^2}{R^2B^2} n^2 \Delta^2 \right\} \right/
\left( \frac{I^2}{R^4}n^2 \Delta^2 + \left( \frac{\vec{B}.\nabla
  B^2}{B^2} \right)^2 \right) 
\end{array}
\end{equation}
Noting that $R^2 \frac{\vec{B}.\nabla B^2}{B^2} \sim B_p^2$ and
$\frac{I^2B_p^2}{R^2B^2} \sim B_p^2$, whereas $\left( \frac{I^2}{R^4} n^2
\Delta^2 + \left( \frac{\vec{B}.\nabla B^2}{B^2} \right)^2 \right)
\sim B^2/R^2$ as $B_p \rightarrow 0$, with $\Delta \sim 1/n$ as found
previously, then we find 
\begin{equation}
| \vec{\xi} |^2 \sim \frac{\left| \xi_{\psi}
  \right|^2}{R^2B_p^2} + \frac{R^2B_p^2}{B^2} \frac{1}{n^2} 
\left| \frac{\partial \xi_{\psi}}{\partial \psi} \right|^2 
\end{equation} 
In other words, we may neglect the term in $\left| \xi_B \right|^2$
compared with the $\left| \xi_{\psi} \right|^2$ and 
$| \xi_{\perp} |^2$ terms. 
We will consider each of these terms in turn. 
Firstly 
\begin{equation}
\begin{array}{ll}
\int \rho_0 \frac{\left| \xi_{\psi} \right|^2}{R^2B_p^2} \vec{dr} &= 
2\pi \int \frac{dl}{B_p} d\psi \rho_0 \frac{|\xi_{\psi}|^2}{R^2B_p^2} 
\\
&\sim  \left. \frac{\rho_0}{R^2} \right|_{\psi_s}\int 
|\xi_{\psi}|^2 d\psi  
 \int \frac{dl}{B_p^3} 
\\
&\lesssim \left. \frac{\rho_0}{I\nabla \phi . \vec{J}}
\right|_{\psi_s} \int q'|\xi_{\psi}|^2 d\psi 
\end{array}
\end{equation}
with $\psi=\psi_s$ at the plasma surface, and 
where we used\cite{Webster08} 
\begin{equation}
\begin{array}{ll}
q'(\psi) &\simeq \frac{1}{2\pi} 
\oint \frac{\nu}{B_p^2} 
\left[ 
\nabla \phi. \vec{J} - \frac{\partial B_p^2}{\partial \psi} 
\right] d\chi 
\\
& \gtrsim 
\frac{\nabla \phi . \vec{J}}{2\pi} 
\oint \frac{dl}{B_p^3} 
\end{array}
\end{equation}
with $dl=(J_{\chi}B_p)d\chi$, and at large aspect ratio
$\nabla \phi . \vec{J}$ is approximately a function of the poloidal
magnetic flux.
Although we have not considered the radial structure of the mode in
this paper, to estimate these terms we will adopt the ansatz that near
the plasma's edge $\xi_m$ can be approximated by a power law, with 
\begin{equation}\label{ximform}
\xi_m = \xi_0 \frac{\left( \psi_a - \psi_s \right)^p}{\left(\psi_a
  -\psi\right)^p} 
\end{equation}
where $\psi=\psi_a$ at the separatrix, and $\psi=\psi_s$ at the plasma
surface. 
This is consistent with studies that do consider a mode's radial
structure\cite{Lortz,Connor}.
We also use the result found here\cite{part2} and
elsewhere\cite{Ryutov}, that for a conventional X-point (as opposed to 
the X-point produced by a ``snowflake'' divertor\cite{Ryutov}),  
near a separatrix we have 
\begin{equation}
q \simeq - q_0 \ln \left( \frac{\psi_a-\psi}{\psi_a} \right) 
\end{equation}
for some constant $q_0\sim 1$. 
Under these assumptions 
\begin{equation}
\int q' | \xi_{\psi} |^2 d\psi \sim |\xi_{\psi}|^2 
\end{equation}
giving 
\begin{equation}
\int \rho_0 \frac{|\xi_{\psi}|^2}{R^2B_p^2}  \vec{dr} 
\sim 
\left. \frac{\rho_0 r }{B\langle B_p \rangle} |\xi_{\psi}|^2
\right|_{\psi_s} 
\end{equation}
where we took $I \nabla \phi \vec{J} \sim B \langle B_p \rangle/r$, 
with $r$ a measure of the plasma radius.

Next we consider, 
\begin{equation}
\int \rho_0 \frac{R^2B_p^2}{B^2} \frac{1}{n^2} 
\left| \frac{\partial \xi_{\psi}}{\partial \psi} \right|^2 \vec{dr} 
\sim \left. \frac{\rho_0}{B^2} \frac{1}{n^2} \right|_{\psi_s} 
\left( \int R^2B_p dl \right) 
\left( \int \left| \frac{\partial \xi_{\psi}}{\partial \psi}
\right|^2 d\psi \right)
\end{equation}
For the trial function
$\xi_{\psi} = \xi_m(\psi) e^{im\theta}$ with $\theta = \frac{1}{q}
\int^{\chi} \nu d\chi'$, we have  
\begin{equation}\label{eqstar}
\int \left| \frac{\partial
  \xi_{\psi}}{\partial \psi} \right|^2  d\psi
\sim \int 
\left( 
\left| \frac{d\xi_m}{d\psi} \right|^2 + \left| \xi_m \right|^2 
\left| \frac{\partial \theta}{\partial \psi} \right|^2 m^2 
\right) d\psi 
\end{equation}
the last expression ignores terms in
$\frac{d\xi_m}{d\psi}\frac{\partial \theta}{\partial \psi}$ because
either $\left|\frac{d\xi_m}{d\psi}\right|$ or $\left| \frac{\partial
  \theta}{\partial \psi} \right|$ will diverge most strongly, so the
largest contribution to the expression will be from either 
$\left|\frac{d\xi_m}{d\psi}\right|^2$ or  $\left| \xi_m\right|^2
\left| \frac{\partial \theta}{\partial \psi} \right|^2m^2$.
Taking $\xi_m$ as in Eq. \ref{ximform}, gives 
\begin{equation}
\int \left| \frac{d\xi_{m}}{d\psi} \right|^2 d\psi 
\sim \left. |\xi_0|^2 q' \right|_{\psi=\psi_s} 
\end{equation}
Similarly, 
because $\oint \nu d\chi = 2\pi q$ and $\theta = \frac{1}{q}
\int^{\chi} \nu d\chi'$, we have 
\begin{equation}
\left| \frac{\partial \theta}{\partial \psi} \right|^2 
= \left[ -\frac{q'}{q}\theta + \frac{1}{q} \int^{\chi}
  \frac{\partial \nu}{\partial \psi} d\chi' \right]^2  
\sim \left( \frac{q'}{q} \right)^2 
\end{equation}
that with $m\simeq nq$ gives 
\begin{equation}
\begin{array}{ll}
\int m^2 |\xi_m|^2 \left| \frac{\partial \theta}{\partial \psi}
\right|^2 
&\sim n^2 \int {q'}^2 |\xi_m|^2 
\\
&\sim \left. n^2 |\xi_0|^2 q' \right|_{\psi=\psi_s} 
\end{array}
\end{equation}
Therefore, because $\psi_s \sim \oint R^2B_pdl/\oint dl$ we have 
\begin{equation}
\int \rho_0 \frac{R^2B_p^2}{B^2} \frac{1}{n^2} 
\left| \frac{\partial \xi_{\psi}}{\partial \psi} \right|^2 \vec{dr} 
\sim \left. \frac{\rho_0 \oint dl}{B^2} \left( \psi_s q'\right) \left|
\xi_{\psi} \right|^2 \right|_{\psi=\psi_s} 
\end{equation}

Finally, taking $\nabla \phi . \vec{J} \neq 0$ and $\Delta' \simeq
-2nq$ ($\Delta'$ is calculated in the second part to this paper),
then gives  
\begin{equation}
\gamma^2 = \frac{-\delta W}{\int \vec{dr} \rho_0 |\xi |^2} 
\sim 
\gamma_A^2 \left( \frac{R_0\vec{J}.\vec{B}}{B^2} \right)^2 
\left( \frac{q}{\psi_s q'} \right) 
\end{equation}
with $\gamma_A \equiv B^2/(\rho_0 R \oint dl)$. 
Because $q'\rightarrow \infty$ more rapidly than $q$ as we approach
the separatrix, then for an outermost flux surface that is made
increasingly close to that  of a separatrix with $\psi_s \rightarrow
\psi_a$, we have that $\gamma^2 
\rightarrow 0$ and  
\begin{equation}
\ln\left(\frac{\gamma}{\gamma_A} \right) = -\frac{1}{2} \ln \left(
\frac{\psi_s q'}{q} \right)  
\end{equation}
with $\gamma_A$ the Alfven frequency, indicating that the 
growth rate $\gamma \rightarrow 0$ as $q'\rightarrow
\infty$.

\section{Summary}

This paper re-explores the stability of the Peeling mode for toroidal
Tokamak geometry.  It starts from a simple approach to Peeling mode
stability at marginal stability  in cylindrical geometry, then
generalises this to toroidal Tokamak equilibrium.   
In the process of doing so we find a number of interesting results, namely 
\begin{enumerate}

\item{At marginal stability, a plasma perturbation induces a skin
  current that is parallel and proportional to the equlibrium current
  at the edge, and proportional to the radial plasma displacement.} 

\item{For zero equilibrium skin current the usual plasma-vacuum
  boundary conditions (Freidberg\cite{Freidberg}), are identical to the
  requirement that $\vec{n}_0 . \vec{B}_1= \vec{n}_0 . \vec{B}_1^V$,
  with the quantities evaluated at the equilibrium position.} 

\item{The equilibrium conditions (force balance) for the Peeling mode
  at marginal stability and high toroidal mode number $n$, are
  identical to requiring $\delta W_S + \delta W_V =0$, where $\delta
  W_S$ and $\delta W_V$ are the surface and vacuum
  contributions to the energy principle's $\delta W =
  \delta W_F + \delta W_S + \delta W_V$, with $\delta W_F$ the
  plasma's contribution to the energy principle
  (Freidberg\cite{Freidberg}). }  

\end{enumerate}
This suggests the Peeling mode be defined as a mode for which $\delta
W_F \ll \delta W_S + \delta W_V$.
For the trial function used by Laval et al\cite{Laval}, that consisted of a
single Fourier mode in straight field line co-ordinates, we find that
the most unstable choice of $\Delta$ gives  
$\delta W = \left( \frac{\pi}{2} \right)^2 \frac{|\xi_m|^2}{R_0}
\left( \frac{2\hat{J}^2}{\Delta'} \right)$.  
To evaluate $\delta W$ for this model, it is necessary to know $\Delta'$
for a plasma cross-section with a separatrix and X-point at the
plasma-vacuum boundary.     
Doing this without making the usual approximations (i.e. with a
discretisation of space), as are usually made in numerical
calculations, is the subject of the second part to this
paper\cite{part2}.   

Finally we considered the growth rate, and found that even with
$\delta W < 0$, the growth rate can be vanishingly small.  
This is because the kinetic energy term was found to diverge like $q'$
as the outermost flux surface becomes increasingly close to a
separatrix.  
When this divergence is sufficiently rapid (as would be the case for
$\Delta'\simeq -2nq$),  then $\ln(\gamma/\gamma_A)$ asymptotes to
$\ln(\gamma/\gamma_A)
= -\frac{1}{2} \ln(\psi_s q'/q)$, a result that may be compared with
those from codes such as ELITE.

{\center{\bf \large Acknowledgments}}

Thanks to Tim Hender for reading and commenting on an earlier draft of
this paper. 
This work was jointly funded  by the United Kingdom Engineering and
Physical Sciences 
Research Council, and by the European Community under the contract of
Association between EURATOM and UKAEA. The views and opinions expressed herein
do not necessarily reflect those of the European commission.

\appendix

\section{Alternative derivation for the growth rate}\label{Altderiv}

Here we provide an alternative derivation for the growth rate, to that
given in Section \ref{grsec}.
In the following we will, 
\begin{enumerate}
\item{Continue to use the high-$n$ ordering of Connor et
  al\cite{Connor}, 
for which $\nabla . \vec{\xi}=0$ requires that 
\begin{equation}
\xi_{\perp} = \frac{i}{n} R^2 B_p^2 \frac{\partial}{\partial \psi}
\xi_{\psi} 
\end{equation}
}\label{as1}
\item{
We will use the arguments from Section \ref{grsec}, that lead us to
expect 
that $\partial \xi_{\psi} / \partial \psi \sim \frac{mq'}{q} |\xi_m| 
\sim nq' |\xi_m|$, for the trial function of Laval et al\cite{Laval}
with 
 $\xi_{\psi}=\xi_m(\psi) e^{im\theta}$. 
}\label{as2}
\item{
As in Section \ref{grsec} we will continue to assume that near the
separatrix, 
\begin{equation}
q \simeq - q_0 \ln \left( \frac{\psi_a - \psi}{\psi_a}
\right) 
\end{equation}
for some constant $q_0 \sim 1$, 
with $\psi_a$ the value of $\psi$ at the separatrix. 
We will also continue to assume that near the separatrix we can
approximate $|\xi_m(\psi)|$ as a power law, with 
\begin{equation}
\xi_m= \xi_0
\frac{(\psi_a-\psi_s)^p}{(\psi_a - \psi)^p}
\end{equation}
where $\psi_s<\ \psi_a$
is the value of $\psi$ at the plasma-vacuum surface. 
}\label{as3}
\end{enumerate}
With the assumptions of \ref{as1}, \ref{as2}, and \ref{as3}, we
require that 
\begin{equation}\label{xiperpa}
\begin{array}{ll}
\xi_{\perp} &= \frac{i}{n} R^2 B_p^2 \frac{\partial
  \xi_{\psi}}{\partial \psi} 
\\
&\sim iR^2B_p^2 q' \left| \xi_m \right| 
\\
&\sim iR^2 B_p^2 \frac{q_0}{\psi_a - \psi} \xi_0 
\frac{(\psi_a - \psi_s)^p}{(\psi_a-\psi)^p} 
\end{array}
\end{equation}
However, we must have $\xi_{\perp} \ll 1$ as $\psi \rightarrow
\psi_s$, and therefore we require $\xi_0 = \hat{\xi}_0 \frac{(\psi_a -
  \psi_s)}{\psi_a}$ with $\hat{\xi}_0\ll 1$ a constant, so that as 
$\psi \rightarrow \psi_a$ we have $\xi_{\perp} \sim \hat{\xi}_0 \ll
1$. 
Therefore we also have as a consequence that 
\begin{equation}
\begin{array}{ll}
\xi_{\psi} &\sim \hat{\xi}_0 \frac{(\psi_a-\psi_s)}{\psi_a}
\frac{(\psi_a-\psi_s)^p}{(\psi_a -\psi)} 
\\
&\rightarrow 
\hat{\xi}_0
\frac{(\psi_a-\psi_s)}{\psi_a} \mbox{ as $\psi \rightarrow \psi_a$}
\end{array}
\end{equation}
In the limit where the plasma surface tends to a separatrix, with
  $\psi_s \rightarrow \psi_a$, we then must have  
$\xi_{\psi} \sim \hat{\xi}_0 \frac{(\psi_a - \psi_s)}{\psi_a}  
  \rightarrow 0$.
Therefore $\delta W$, for which $\xi_m$ is evaluated at $\psi=\psi_s$,
  has 
\begin{equation}\label{aa6}
\begin{array}{ll}
\delta W &\sim - \frac{\left| \hat{\xi}_0 \right|^2}{R_0}
\left( \frac{R_0 \vec{J} . \vec{B}}{B^2} \right) 
\left( \frac{q}{n} \right) 
\left( \frac{\psi_a -\psi_s}{\psi_a} \right)^2
\\
& \rightarrow 0 \mbox{ as $\psi_s \rightarrow \psi_a$}
\end{array}
\end{equation}
Next we consider the growth rate.

Using arguments from Section \ref{grsec}, we expect 
\begin{equation}
\int \vec{dr} \left| \xi \right|^2 \sim \int \vec{dr} 
\frac{\left| \xi_{\perp} \right|^2}{R^2B_p^2 B^2} 
\end{equation}
Now with $\xi_{\perp}$ given by Eq. \ref{xiperpa}, we have 
\begin{equation}\label{aa8}
\begin{array}{ll}
\int \vec{dr} 
\frac{\left| \xi_{\perp} \right|^2}{R^2B_p^2 B^2} 
&\sim 
\int \frac{dl d\psi d\phi}{B_p} 
\frac{R^2B_p^2}{B^2} 
\left( \frac{\psi_a-\psi_s}{\psi_a - \psi} \right)^{p+1} 
\frac{\left|\hat{\xi}_0\right|^2}{\psi_a^2}
\\
&\sim \left| \hat{\xi}_0 \right|^2 \frac{R^2}{\psi_a^2} 
\left( \oint \frac{B_p dl}{B^2} \right) 
\int d\psi \left( \frac{\psi_a-\psi_s}{\psi_a - \psi} \right)^{p+1}  
\\
&\sim  \frac{\left| \hat{\xi}_0 \right|^2}{p} 
\frac{\oint dl}{\langle B^2 \rangle} 
\left( \frac{\psi_a-\psi_s}{\psi_a} \right) 
\end{array}
\end{equation}
where $\langle \rangle$ denotes a poloidal average, and 
in the last line we used $\psi_a \sim \langle B_p \rangle R^2$. 
Hence using Eqs. \ref{aa6} and \ref{aa8} we find 
\begin{equation}
\begin{array}{ll}
\gamma^2 &= \frac{\delta W}{\int \vec{dr} \rho_0 |\xi |^2 }
\\
& \sim 
\left.
- \frac{\left| \hat{\xi_0} \right|^2}{R_0}
\left( \frac{R_0 \vec{J} . \vec{B}}{B^2} \right) 
\left( \frac{q}{n} \right) 
\left( \frac{\psi_a -\psi_s}{\psi_a} \right)^2 
\right/
\frac{\left| \hat{\xi}_0 \right|^2}{p} 
\frac{\oint dl}{\langle B^2 \rangle} 
\left( \frac{\psi_a-\psi_s}{\psi_a} \right) 
\\
&\sim \gamma_A^2 
\left( \frac{R_0 \vec{J} . \vec{B}}{B^2} \right) 
\left( \frac{\psi_a-\psi_s}{\psi_a} \right) \frac{pq}{n} 
\end{array}
\end{equation}
with $\gamma_A^2 = \langle B^2 \rangle/\rho_0 R_0 \oint dl$. 
Therefore as the outermost plasma surface more closely approximates a
separatrix with $\psi_s \rightarrow \psi_a$, we have that
$\gamma/\gamma_A \rightarrow 0$. 
Note that because we have taken $q\sim q_0 \ln
\left(\frac{\psi_a-\psi}{\psi_a} \right)$, then $q' \sim q_0/(\psi_a -
\psi)$, and hence $\left( \gamma/\gamma_A \right)^2 \sim 1/q'$ as found 
Section \ref{grsec}.


\end{document}